\documentclass[twocolumn,showpacs]{revtex4}
\newcommand{\gtsim}{\mbox{{\raisebox{-0.4ex}{$\stackrel{>}{{\scriptstyle\sim}}$}}}}
\newcommand{\ltsim}{\mbox{{\raisebox{-0.4ex}{$\stackrel{<}{{\scriptstyle\sim}}$}}}}

\usepackage{graphicx}
\usepackage{dcolumn}
\usepackage{amsmath}
\begin{document}

\title{Landau quantization effects in the charge-density-wave system
(Per)$_2M$(mnt)$_2$ (where $M=$~Au and Pt)}

\author{R.~D.~McDonald$^1$, N.~Harrison$^1$, 
J.~Singleton$^1$, A.~Bangura$^2$, P.A. Goddard$^1$,
A.~P.~Ramirez$^3$ and X.~Chi$^3$}

\affiliation{$^1$National High Magnetic Field Laboratory, LANL, MS-E536, Los Alamos, New Mexico 87545\\
$^2$The Clarendon Laboratory, Parks Road, Oxford OX1 3PU, United Kingdom\\
$^3$Bell Laboratories, Lucent Technologies, 600 Mountain Avenue, Murray Hill, 
New Jersey 07974}
\date{\today}

\begin{abstract}
A finite transfer integral $t_a$ orthogonal to the
conducting chains of a highly one-dimensional 
metal gives rise to
empty and filled bands
that simulate an indirect-gap semiconductor
upon formation of a commensurate charge-density-wave (CDW).
In contrast to semiconductors such as Ge and Si
with bandgaps $\sim 1$~eV, the CDW system
possesses an indirect gap with a greatly reduced energy scale,
enabling moderate laboratory magnetic fields to have 
a major effect. The consequent variation of the thermodynamic
gap with magnetic field due to Zeeman
splitting and Landau quantization enables the electronic 
bandstructure
parameters (transfer integrals, Fermi velocity)
to be determined accurately. 
These parameters reveal the orbital
quantization limit to be reached at $\sim 20$~T in
(Per)$_2M$(mnt)$_2$ salts, making them highly unlikely candidates for a
recently-proposed cascade of field-induced charge-density wave states.
\end{abstract}
\maketitle 

\begin{figure}[htbp]
\centering
\includegraphics[width=8.5cm]{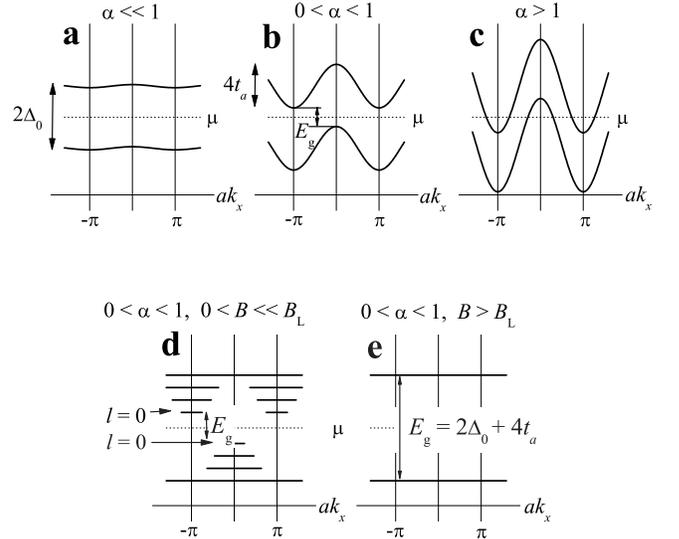}
\caption{Electronic dispersion at $k^\prime_y=$~0 according to
Eq.~\ref{gapdispersion} for differing values of
$\alpha=2t_a/\Delta$; {\bf a}:~$\alpha\ll$~1, 
{\bf b}:~0~$<\alpha<$~1, {\bf c}:~$\alpha>$~1. 
Scenario {\bf b} corresponds to
an indirect gap semiconductor.  
{\bf d}: Schematic of the Landau quantization of scenario
{\bf b} in the presence 
of finite $B$ (ignoring the Zeeman effect, for clarity).
{\bf e}: The gap beyond the field limit
for Landau quantization ($B>B_{\rm L}$; see text).
In all diagrams, $\mu$ represents the chemical potential.}
\label{diagram}
\end{figure}

Magnetic fields $B$ affect the energies of band electrons via
Landau (orbital) quantization and Zeeman (spin) splitting~\cite{ashcroft1}.
In semiconductors such as GaAs and Si, the characteristic
energies of these effects for $B~\ltsim~100$~T 
are much smaller than the energy gap, $E_{\rm g}$,
and the conduction- and valence-band widths, all of which are
$\sim 1$~eV;
hence, the field can be regarded as a 
mere perturbation of the overall bandstructure. 
By contrast, commensurate charge-density wave (CDW) 
groundstates have the potential to provide analogues in which
$E_{\rm g}$ and the bandwidths are scaled down by factors
$\sim 10-1000$~\cite{gruner1}. An ``indirect-gap
semiconductor'' is produced in a CDW
if the characteristic bandwidth $4t_a$
of the electronic dispersion orthogonal to the chain direction (and
nesting vector ${\bf Q}$) is comparable to but smaller than the
order parameter $2\Delta$ (see Fig.~\ref{diagram} {\bf b}).
In such circumstances, the small size of $E_{\rm g}$ means
that the Landau and Zeeman  energies due to typical
laboratory fields are major effects, resulting in
marked alterations to the overall electronic properties.

In this Letter, we show that CDW materials of the composition
(Per)$_2M$(mnt)$_2$, with $M=$~Au and Pt, are ideal candidates for
this purpose~\cite{henriques1}.  Fields of
$B<45$~T have been shown to suppress the CDW groundstates
of these systems~\cite{graf1,mcdonald1,graf2,mcdonald2}, 
providing an estimate for $2\Delta$ of a few meV, 
only slightly larger than the $4t_a \sim 1$~meV obtained from 
bandstructure calculations~\cite{veiros1}.  
Landau quantization and Zeeman splitting
of the filled and
empty states (Fig.~\ref{diagram}{\bf d}) leads to
a field-dependent thermodynamic energy gap 
$E_{\rm g}(B)$, that may, with care, 
be extracted from the thermally-activated component 
of the conductivity
\begin{equation}\label{activation}    
\sigma_T=\sigma_0\exp[-E_{\rm g}(B)/2k_{\rm B}T],
\end{equation}
where $T$ is the temperature.
The measured $E_{\rm g}(B)$ 
provides a means of deducing both $4t_a$ and the 
Fermi velocity ${\bf v}_{\rm F}$ 
along the chains. These values
are consistent with bandstructure calculations~\cite{veiros1}
and are in good agreement with
estimates from thermopower data~\cite{henriques1}.
The data also enable us to identify the maximum field at which
closed orbits can exist, which is found to be too small to support
recently-postulated field-induced charge-density wave states
in (Per)$_2M$(mnt)$_2$ salts~\cite{graf2}.

Charge transfer between (Per)$^+_2$ and $M$(mnt)$^-_2$ in
(Per)$_2M$(mnt)$_2$ gives rise to a normal metallic state with a
highly one-dimensional (1D) $3/4$-filled band in which 
${\bf v}_{\rm F}$ is directed along the Perylene chains 
parallel to the
crystallographic ${\bf b}$ axis~\cite{veiros1}.  
Commensurate CDW groundstates occur for $M=$~Pt, 
Cu and Au, in which the 1D band becomes gapped upon 
modulation of the lattice at a 
postulated~\cite{gama1}
wavevector ${\bf Q}=(0,2k_{\rm F},0)$, where 
$\pm k_{\rm F}=\frac{\pi}{4}|{\bf b}|$~\cite{lopes1}.  
Transitions into the CDW groundstate occur at
$T_{\rm P}\approx 12$~K for $M=$~Au and 
$T_{\rm P}\approx 8$~K for $M=$~Pt.  
In the case of $M=$~Pt, this is accompanied by the
synchronous dimerization of the $S=1/2$ Pt spins~\cite{lopes1}: 
we return to the special case of $M=$~Pt below. 
In all cases, semiconducting behaviour occurs 
for $T \leq T_{\rm P}$; as $T$ is reduced further,
in addition to the thermally-activated conduction,
there is an increasing contribution from the sliding 
collective mode of the CDW as a characteristic
threshold electric field ${\cal E}_{\rm t}$
is approached~\cite{mcdonald1}.
\begin{figure}[htbp]
\centering
\includegraphics[width=8.5cm]{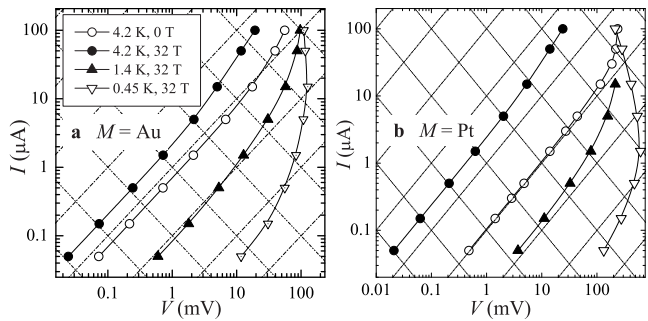}
\caption{Non-linear current-versus-voltage characteristics of
(Per)$_2$Au(mnt)$_2$ ({\bf a})
and (Per)$_2$Pt(mnt)$_2$ ({\bf b})
plotted on logarithmic scales for various
temperatures and fields (see inset key).
The negative-slope diagonal lines are contours 
of constant power and the 
positive-slope diagonal lines are contours of 
constant resistance, providing a guide 
as to when the sample's behavior is dominated by ohmic,
thermally-activated conduction rather than sliding.}
\label{IV}
\end{figure}

In these experiments, $E_{\rm g}(B)$
is determined for $M=$~Au and Pt using transport
measurements in static fields of up to 33~T, with
$0.45~{\rm K} \leq T \leq 4.2$~K provided by a $^3$He cryostat.  
Currents as low as $I=50$~nA (equivalent to a current
density $j_y\approx 50$~Am$^{-2}$) applied
along the crystal's $b$ direction, enable $E_{\rm g}$ to be
extracted from the $T$-dependence of the
resistance $R$ (proportional to the resistivity tensor
component $\rho_{yy}$~\cite{mcdonald2}) 
over a range of $T$ where the contribution from the 
CDW collective mode is small.  Figure~\ref{IV} shows 
that this is the case over a wide range of $B$ provided 
$I~\ltsim~100$~nA and $T~\gtsim~1$~K.

\begin{figure}[htbp]
\centering
\includegraphics[width=8.5cm]{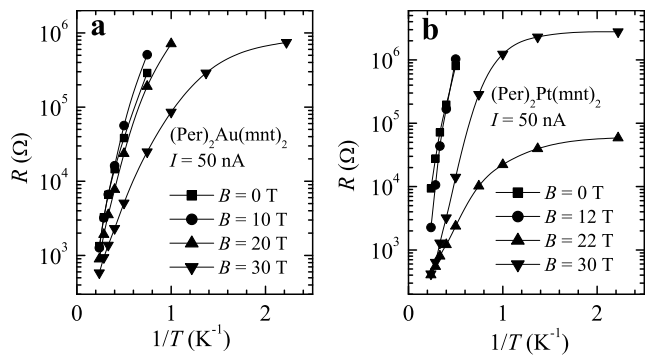}
\caption{Arrhenius plots of $R$ (logarithmic scale)
versus $1/T$ with $I=$~50~nA 
for ({\bf a})~$M=$~Au and ({\bf b})~$M=$~Pt 
at several different magnetic field strengths $B$.}
\label{arrhenius}
\end{figure}

Figures \ref{arrhenius}{\bf a} and {\bf b} show examples of 
Arrhenius plots of the resistance $R$ versus reciprocal 
temperature $1/T$ for
$M=$~Au and Pt respectively, 
made using $I=$~50~nA with 
${\bf B}||{\bf c}^\ast$ at several different $B$.  
Prior studies~\cite{mcdonald1,mcdonald2}
have shown that $\rho_{yy}$ in (Per)$_2M$(mnt)$_2$ conforms to
a simple parallel conduction model in which the Hall component of the
conductivity tensor $\sigma_{xy}\approx 0$, leading to
\begin{equation}\label{resistivity}
\rho_{yy}=    (\sigma_T+j_y/{\cal E}_{\rm t})^{-1},
\end{equation}
where $j_y/{\cal E}_{\rm t}$ is the contribution from the collective mode. 
As discussed below, the expression for $E_{\rm g}(B)$ (Eq.~\ref{gapfield}) 
contains $\Delta$, which is $T$-dependent;
moreover, ${\cal E}_{\rm t}$ may depend on $T$.
Thus, the slope of an Arrhenius plot becomes 
\begin{equation}\label{slope}
\frac{\partial\ln\rho_{yy}}{\partial (1/T)} = 
\frac{\frac{1}{2k_{\rm B}}(E_{\rm g}-T\frac{\partial E_{\rm  g}}{\partial T})
-\frac{2k_{\rm B}j_yT^2}{\sigma_T{\cal E}_{\rm t}^2}    
\frac{\partial {\cal E}_{\rm t}}{\partial T}}
{1+\frac{j_y}{\sigma_T {\cal E}_{\rm  t}}}.
\end{equation}
Since ${\cal E}_{\rm t}$ is expected to
depend strongly on $T$ only as 
$T\rightarrow T_{\rm P}$~\cite{maki1}, 
the last term on the right-hand-side of Eq.~\ref{slope} 
should be negligible for $T\ltsim~T_{\rm P}/2$ and
sufficiently small $j_y$. The second term can also be minimized by
evaluating the slope for $T<T_{\rm P}/2$; using a mean-field
expression for $\Delta(T)$~\cite{gruner1}, these conditions give 
$-2T\partial E_{\rm g}/\partial T \equiv -T\partial\Delta/\partial T<0.1\Delta$.  
In the present case, we choose $T\approx T_{\rm P}/3$, 
such that
$-T\partial\Delta/\partial T\approx$~0.05$\Delta$, yielding
$\partial\ln(\rho_{yy})/\partial(1/T)\approx E_{\rm g}/2k_{\rm B}$ with a
systematic error of only $\sim 5$~\%. 
In using this method to extract $E_{\rm g}$,
it is important to 
note that $T_{\rm P}=T_{\rm P}(B)$~\cite{graf1}; thus, 
the exact $T$
range used depends on the applied field.

At lower $T$, the
number of thermally-activated carriers decreases significantly,
rendering their contribution to the conductivity negligible. 
The consequent larger potential difference across the sample
results in increased depinning of the CDW, 
so that $\ln(R)$ saturates at lower
$T$ (i.e. higher $1/T$) in Fig.~\ref{arrhenius}.  
As shown in Fig.~\ref{IV}, the point at which significant 
CDW depinning occurs is shifted to lower $T$ (higher $1/T$)
by the use of small currents.

By using larger currents of 1 and $10~\mu$A, 
Graf {\it et al}~\cite{graf1} obtained an Arrhenius plot in 
which the CDW was highly depinned over an extensive range of 
$T$, causing $\ln(\rho_{yy})$ versus $1/T$ to exhibit 
significant curvature even at 
$T\sim T_{\rm P}/3$. To make matters worse, Graf {\it et al.} 
extracted $\partial \rho_{yy}/\partial (1/T)$
at $T\approx$~0.7~$T_{\rm P}$, at which $-T\partial\Delta/\partial T>\Delta$.
Sadly, these factors introduced 
a systematic error $\sim 100$~\%, 
resulting in an overestimate of $E_{\rm g}$ by
a factor $\sim2$.

\begin{figure}[htbp]
\centering
\includegraphics[width=8.5cm]{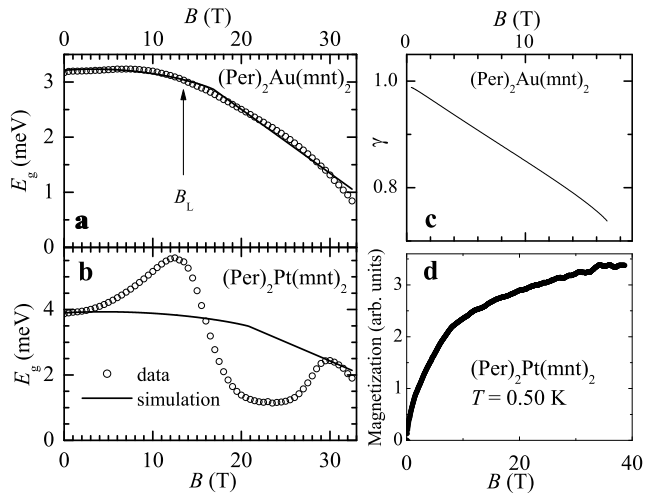}
\caption{The excitation gap $E_{\rm g}$ (circles) 
for $M=$~Au ({\bf a}) and $M=$~Pt ({\bf b})
estimated from the slopes of Arrhenius
plots like those in Fig.~\ref{arrhenius}
at $T\approx T_{\rm P}/3$ at many different values of 
the applied $B$. The solid lines represent fits of 
the model described in the text.
({\bf c})~The parameter 
$\gamma(B)$ obtained by fitting the 
experimentally-estimated gap in {\bf a}.
({\bf d})~Magnetization of many randomly-orientated
$M=$~Pt crystals at $T=0.50$~K measured using an
extraction magnetometer in a pulsed magnetic 
field~\cite{goddard1}.}
\label{gap}
\end{figure}

The present estimate of $E_{\rm g}\approx$~3.2~meV 
made at $T\approx T_{\rm p}/3$ and $B=0$ with $I=$~50~nA, 
shown in Fig.~\ref{gap}{\bf a}, is rather less 
than the value of $2\Delta_0 \approx 4.5$~meV deduced
for $M=$~Au from its Pauli paramagnetic
limit~\cite{mcdonald1},
supporting the scenario
depicted in Fig.~\ref{diagram}{\bf b}.
This can be modeled using a simplified dispersion of the form
\begin{equation}
\label{dispersion}
\varepsilon=-2t_a\cos ak_x-\hbar v_{\rm F}|k_y-k_{\rm F}|,
\end{equation}
where $v_{\rm F}=\sqrt{2}bt_b/\hbar$ is the magnitude of 
${\bf v}_{\rm F}$ directed along ${\bf b}$ for a $\frac{3}{4}$-filled band.  
Here, $t_a$, $t_b$, $a (\approx 17$~\AA) and $b (\approx 4.2$~\AA) are
the transfer integrals and lattice spacings perpendicular and parallel
to the chains respectively.
On opening a gap $2\Delta$ (which
may be smaller than $2\Delta_0$ at finite $T$), 
the dispersion becomes~\cite{maki1}
\begin{equation}
\label{gapdispersion}
\varepsilon_{\rm g}=-2t_a\cos ak_x \pm \sqrt{(\hbar v_{\rm F}k^\prime_y)^2+\Delta^2},
\end{equation}
where we have substituted $k^\prime_y=k_y-k_{\rm F}$ for clarity.  
At $k^\prime_y=$~0, the electronic dispersion versus $k_x$
consists of a pair of narrow bands above and below
$\mu$, separated in energy by $2\Delta$, as shown in
Fig.~\ref{diagram} for different values of 
$\alpha=2t_a/\Delta$.  
When $\alpha\ll  1$, one obtains a gap with little dispersion.  
This is the likely scenario for CDW materials with very large order 
parameters such as (TaSe$_4$)$_2$I or TaS$_3$~\cite{gruner1}.
In the opposite extreme, $\alpha>1$, one
obtains a semimetal qualitatively similar to NbSe$_3$~\cite{gruner1}
or perhaps $\alpha$-(BEDT-TTF)$_2M$Hg(SCN)$_2$, in which field-induced
CDW phases have been suggested~\cite{andres1}.
An indirect gap semiconductor with $E_{\rm g}=2\Delta-4t_b$ 
between filled and empty states separated by the vector 
$[\pi/a,0]$ is obtained for the
intermediate situation where $0<\alpha< 1$.

The introduction of finite $B$ has two effects on Eq.~\ref{gapdispersion}.  
The first is to Zeeman split the two bands into four subbands with energies
\begin{equation}
\label{Zeemandispersion}
\varepsilon_{\rm g}=-2t_a\cos ak_x \pm \sqrt{(\hbar v_{\rm F}k^\prime_y)^2
+\Delta^2}+gs\mu_{\rm B}B,
\end{equation}
where $s~(=\pm\frac{1}{2})$ is the electron spin and 
$g~(\approx 2)$ is its Land\'{e} g-factor.
Thus, the smallest ({\it thermodynamic}) gap
occurs between the maximum of the $s=+\frac{1}{2}$ subband
below $\mu$ and the minimum of the $s=-\frac{1}{2}$
subband above $\mu$,
yielding $E_{\rm g}=2\Delta-4t_a-g\mu_{\rm B}B$.
The second effect of finite $B$ is
to quantize the orbital motion of the empty and filled states
immediately above and below the gap, giving rise to sets of completely
empty and filled Landau levels at $T=0$ (Fig.~\ref{diagram}{\bf d}).  
At finite $T$, carrier excitations
occur between the $l=0$ Landau levels on either side of $\mu$,
leading to a thermodynamic gap
\begin{equation}
\label{gapfield}
E_{\rm g}(B)=2\Delta-4t_a-g\mu_{\rm B}B+\gamma\hbar\omega_{\rm c}
\end{equation}
where $\omega_{\rm c}=eB/m^\ast$ is the magnitude of
the cyclotron frequency of the
$l=0$ Landau level of the two relevant subbands
in the limit $B\rightarrow 0$ and $m^*$ is an
effective mass, to be defined below.

The parameter $\gamma$ describes the (identical)
nonparabolicities of the $s=+\frac{1}{2}$ subband
below $\mu$ and the $s=-\frac{1}{2}$
subband above $\mu$; it is derived by equating the
$k$-space area $A_k(\varepsilon_{\rm g})$
of an orbit of constant energy 
$\varepsilon_g$ (from Eq.~\ref{Zeemandispersion})
with the Onsager $k$-space area $A_{l=0}=\pi eB/\hbar$ of
the $l=0$ Landau level~\cite{ashcroft1}.
This cannot be done analytically;
instead, the fit (solid line) to the
data in Fig.~\ref{gap}{\bf a} for $M=$~Au
incorporates a numerical integration
$A_k(\varepsilon_{\rm g})=\int\Theta|\varepsilon_{\rm g}|{\rm d}k_x{\rm d}k^\prime_y$,
where $\Theta$ is the theta function
({\it i.e.} the integral of the Kronecker $\delta$).

The fit involves adjusting three parameters, $2\Delta$, $4t_a$
and $v_{\rm F}$.  A satisfactory fit is obtained for
$M=$~Au using $4t_a=$~0.80~$\pm$~0.03~meV, 
$v_{\rm F}=$~(1.70~$\pm$~0.05)~$\times 10^5$~ms$^{-1}$ and 
$2\Delta =$~4.02~$\pm$~0.03~meV;
note that over the $T$ range used, the deduced
value of $\Delta$ is equal to $\Delta_0$ to a good approximation.
These parameters correspond to $E_{\rm g}=$~3.21~$\pm$~0.03~meV 
at $B=0$, $T=0$ and an intrachain bandwith of $4t_b\approx$~752~meV;
the latter value is very close to estimates
from thermopower data
($4t_a\approx 740$~meV)~\cite{henriques1}
and in reasonable accord with bandstructure
calculations ($4t_a \approx 590$~meV)~\cite{veiros1}.

Figure~\ref{gap}{\bf c} shows the
field dependence of
$\gamma$ in Eq.~\ref{gapfield}.  
At $B=$~0, $\gamma= 1$ due to the approximately 
parabolic curvature of
the bands close to their extrema
\begin{equation}
\label{approxdispersion}
\varepsilon_{\rm g}\approx 2\Delta-4t_a\pm g\mu_{\rm B}B
+\frac{\hbar^2k^2_x}{2m_a}+ \frac{\hbar^2k^{\prime2}_y}{2m_\Delta}+\ldots
\end{equation}
Here, $m^\ast=0.90m_{\rm e}$ is an effective
mass ($m_{\rm e}$ is the free electron mass), given by
$m^\ast=\sqrt{m_am_\Delta}$ for an approximately elliptical orbit
where $m_a=\hbar^2/2a^2t_a\approx$~50~$m_{\rm e}$ and
$m_\Delta=\Delta/v^2_{\rm F}\approx$~0.01~$m_{\rm e}$.  
Eq.~\ref{gapfield} therefore describes a situation in which 
$E_{\rm g}$ initially increases with $B$ at a rate 
$\partial E_{\rm g}/\partial B\approx
\hbar e(m_{\rm e}-m^\ast)/m^\ast m_{\rm e}$, owing to
the fact that $m^\ast<m_{\rm e}$ and $g\mu_{\rm B}\approx \hbar e/m_{\rm e}$.  
As $B$ increases further, the nonparabolic curvature of 
Eq.\ref{Zeemandispersion} at 
$|\varepsilon_{\rm g}|>\Delta-2t_a-gs\mu_{\rm B}B$ 
eventually causes $E_{\rm g}$ to fall, corresponding to a reduction 
in $\gamma$ in Fig.~\ref{gap} (inset).
This occurs until the $l=0$ 
Landau level of each subband acquires an energy of magnitude
$|\varepsilon_{\rm g}(B)|=|(\Delta+2t_a-gs\mu_{\rm B}B_{\rm L})|$ 
at the limit 
\begin{equation}
\label{quantumlimit}
B_{\rm L}=8t_am^\ast/\gamma\hbar e,
\end{equation}
at which $\gamma\hbar\omega_{\rm c}=8t_a$, $k_x=\pi/a$ and closed
orbits can no longer exist.  This limit corresponds to 
the largest possible pocket created by imperfect nesting matching the
$k$-space area $A_{l=0}\approx$~10$^{17}$~m$^{-1}$
($\approx 0.15$~\% of the Brillouin zone) of the $l=0$ Landau level.  
Orbitally-quantized states can therefore not exist in
(Per)$_2$Au(mnt)$_2$ at fields greater than 17~T, supporting the
existence of an inhomogeneous CDW phase in this material at
$B\gtrsim 30$~T~\cite{mcdonald2} as opposed to a cascade of
field-induced CDW states involving orbital quantization~\cite{andres1,lebed1}.
The absence of Landau subbands
for $B\geq B_{\rm L}\approx$~17~T ($M$=Au)
leads to a simpler field-dependent gap
\begin{equation}
\label{highfield}
E_{\rm g}=2\Delta+4t_a-g\mu_{\rm B}B.
\end{equation}
The slope for $B>B_{\rm L}$ is determined entirely by $|gs|\approx$~1,
making it independent of all fitting parameters.  
Only the intercept of Eq.~\ref{highfield} at $B=0$ 
is determined by $2\Delta+4t_a$.
Hence, the portion of the fit for $B>B_{\rm L}$ in
Fig.~\ref{gap}{\bf a} effectively involves the adjustment of only a
single parameter, imposing a severe contraint on $2\Delta+4t_a$ and
thereby adding further confidence in the model.

For $M=$~Au, we have shown that
a uniform CDW order parameter
$2\Delta_0\approx 2\Delta =4.02\pm0.03$~meV,
bandwidth $4t_a=$~0.80~$\pm$~0.03~meV and
Fermi velocity 
$v_{\rm F}=$~(1.70~$\pm$~0.05)~$\times 10^5$~ms$^{-1}$
explain the experimentally-observed
field-dependence of $E_{\rm g}$.  
This requires a revision of our
estimate for the Pauli paramagnetic limit to 
$B_{\rm P}=(\Delta_0+2t_a)/\sqrt{2}gs\mu_{\rm B}\approx 30$~T, which now
approximately corresponds to the field at 
which ${\cal E}_{\rm t}$ starts to
drop experimentally~\cite{mcdonald1}.
The revised estimate of $2\Delta_0$ further yields 
$\zeta=k_{\rm B}T_{\rm P}/2\Delta_0\approx$~3.9: 
since this estimate only marginally exceeds the BCS value 
of $\zeta=$~3.52, it implies that the CDW groundstate of
(Per)$_2M$(mnt)$_2$ with $M=$~Au is weakly coupled to the crystalline
lattice~\cite{gruner1}.

Turning now to the case of $M=$~Pt, it is clear that $E_{\rm g}$ is
larger at $B=$~0, in spite of the lower $T_{\rm p}(\approx 8$~K). The
complex field-dependent behavior that takes place for $3<B<30$~T 
only in the case of $M=$~Pt must be the consequence of
interactions involving the dimerization of the $S=1/2$ Pt spins, which
are not included in the present simulation.
In situations in which the magnetization
(Fig.~\ref{gap}{\bf d})
is either small ($B\approx 0$)
or approaching saturation ($B\approx 30$~T), 
the use of $4t_a=1$~meV and the same $v_{\rm F}$ as in $M=$Au
results in a fitted value $2\Delta\approx 4.8$~meV.
In the absence of a more sophisticated model, one can
conclude from the estimate of $\zeta=2\Delta_0/k_{\rm B}T_{\rm P}\approx$~7, 
that the dimerization of the Pt spins in (Per)$_2$Pt(mnt)$_2$ 
causes the CDW groundstate to be more strongly coupled to the 
crystalline lattice than in the case of $M=$Au.
The consequent larger lattice distortion in $M=$~Pt
might explain its higher ${\cal E}_{\rm t}$ values
compared to those of  $M=$~Au (Fig.~\ref{IV}).
It may also be the reason why
X-ray Bragg reflection peaks 
due to dimerization have been seen
for $M=$~Pt but not for $M=$~Au~\cite{henriques1}.

In conclusion, (Per)$_2$Au(mnt)$_2$ is shown to possess an ideal
combination of parameters for modeling the effect of a strong magnetic
field on an indirect gap semiconductor.  
A model that includes the effect of Zeeman splitting 
and Landau quantization of subbands with a conventional 
field-independent order parameter (for fields below
the Pauli paramagnetic limit)
provides a good description of the experimental data~\cite{note}.
The electronic structure parameters obtained reveal the
limiting magnetic field
for closed orbits to be $B_{\rm L}\approx$~17~T, implying
field-induced CDW states that incorporate orbitally quantized levels
cannot exist at $B\gtrsim$~17~T in (Per)$_2$Au(mnt)$_2$.  The
electronic structure of (Per)$_2$Pt(mnt)$_2$ is expected to be
similar.  However, the existence of $S=1/2$ Pt spins causes a stronger
coupling of the CDW to the lattice followed by a more complex dependence of
$E_{\rm g}$ on field for $B\lesssim 30$~T that has yet to be modeled.

This work is supported by US Department of Energy (DOE) grants LDRD20030084DR 
and LDRD2004009ER and was performed under the auspices 
of the National Science Foundation, the DOE and the State of Florida.
We thank Chuck Mielke and 
Albert Migliori for very useful comments.

\end{document}